\documentclass[runningheads]{llncs}
\usepackage[T1]{fontenc}
\usepackage{graphicx}
\usepackage{amsmath}
\usepackage{graphicx, float, placeins}
\usepackage{esvect}

\begin{document}

\title{\textsc{gr-orbit-toolkit}: A Python-Based Software for Simulating and Visualizing Relativistic Orbits}


\author{Milagros Delgado\inst{1}\orcidID{0009-0001-8391-4398} \and\\
Wladimir E. Banda-Barrag\'{a}n\inst{1}\orcidID{0000-0002-1960-4870}}

\authorrunning{M. Delgado and W. E. Banda-Barrag\'{a}n}

\institute{Escuela de Ciencias F\'isicas y Nanotecnolog\'ia, Universidad Yachay Tech, Hacienda San Jos\'e S/N, 100119 Urcuqu\'i, Ecuador\\
\email{\{milagros.delgado,wbanda\}@yachaytech.edu.ec}\\
\url{https://yachaytech.edu.ec/en/escuela-de-ciencias-fisicas-y-nanotecnologia}}

\titlerunning{\textsc{gr-orbit-toolkit}: Simulating Relativistic Orbits}
%
\maketitle              
\begin{abstract}
Creating software dedicated to simulation is essential for teaching and research in Science, Technology, Engineering, and Mathematics (STEM). Physics lecturing can be more effective when digital twins are used to accompany theory classes. Research in physics has greatly benefited from the advent of modern, high-level programming languages, which facilitate the implementation of user-friendly code. Here, we report our own Python-based software, the \textsc{gr-orbit-toolkit}, to simulate orbits in classical and general relativistic scenarios. First, we present the ordinary differential equations (ODEs) for classical and relativistic orbital accelerations. For the latter, we follow a post-Newtonian approach. Second, we describe our algorithm, which numerically integrates these ODEs to simulate the orbits of small-sized objects orbiting around massive bodies by using Euler and Runge-Kutta methods. Then, we study a set of sample two-body models with either the Sun or a black hole in the center. Our simulations confirm that the orbital motions predicted by classical and relativistic ODEs drastically differ for bodies near the Schwarzschild radius of the central massive body. Classical mechanics explains the orbital motion of objects far away from a central massive body, but general relativity is required to study objects moving at close proximity to a massive body, where the gravitational field is strong. Our study on objects with different eccentricities confirms that our code captures relativistic orbital precession. Our convergence analysis shows the toolkit is numerically robust. Our \textsc{gr-orbit-toolkit} aims at facilitating teaching and research in general relativity, so a comprehensive user and developer guide is provided in the public code repository.

\keywords{Open-source Software \and STEM Education Tools \and Digital Twin Software Platforms \and General Relativity \and Scientific Visualization}
\end{abstract}

\section{Introduction}
The study of the orbital motion of bodies moving under the influence of a strong gravitational potential of a massive central object requires General Relativity (GR). While Newtonian mechanics assumes gravitational forces regulate such interactions via an inverse square law \cite{Celletti_2010}, GR links gravity to space-time curvature. While Newton's interpretation is intuitive and works for weak gravitational potentials (e.g., Earth's orbit around the Sun), relativistic corrections are required when studying orbits near massive bodies (e.g., Mercury's orbit around the Sun \cite{Hartle_2021}).\par

Despite the success of GR at explaining astrophysical observations, such as the orbital precession of Mercury \cite{Park_2017}, the existence of black holes and gravitational waves \cite{Abbott_2016}, light bending and gravitational lensing \cite{Will_2014}, and the cosmological expansion of the Universe \cite{Ishak_2018}, the mathematical theory of GR is highly non-linear. Solving the GR equations often requires simplifying the mathematics under specific assumptions to obtain non-linear ODEs and the use of numerical approximation methods to find particular solutions. The solutions are sensitive to the initial conditions of the system, and chaos can readily emerge \cite{Kovács_2011,Portegies_2022}. Finding stable orbits around massive objects can then become a complex mathematical problem, and computational tools are required (e.g., see \cite{Baumgarte_2010}, \cite{Blackman_2017}, and \cite{Buonanno_1999}).\par

With the advent of modern computing and the development of increasingly faster computers, creating specialized simulation software has become accessible. High-level programming languages, like Python, are ideal for implementing simulation software as they provide user-friendliness, portability, and flexibility. Python-based software can also be readily combined with other languages for optimization and converted to JavaScript code for integration with HTML-based websites (e.g., see the PhET Interactive Simulation Project \cite{2008AmJPh..76..406M}). Some examples of software developed for simulating orbital motion include: ODTBX \cite{2018ascl.soft10010N} (which is Matlab-based), AWS Orbit Workbench\footnote{https://awslabs.github.io/aws-orbit-workbench} (which is Python-based), Fourmilab's applet\footnote{https://www.fourmilab.ch/gravitation/orbits} (which is JavaScript-based), among others. Some of these tools, and other more specialized software (such as the N-body simulation package REBOUND \cite{2012A&A...537A.128R} and the digital applet GeodesicViewer \cite{2010CoPhC.181..413M}) have been developed with both research and teaching purposes. These packages can assist lecturers in teaching orbital mechanics and relativity.\par

The use of STEM education tools, like digital applets, in physics curricula is becoming increasingly popular. Physlets, for instance, has been used since the early 2000s for physics teaching \cite{Christian_2020}. The Physlets database contains hundreds of JavaScript-based applets that allow users to interact with scroll bars and set customized conditions to perform simulations. GeoGebra has also been extensively used for mathematics teaching, particularly geometry, as reported in \cite{Rodríguez_2013}. Of course, applets and technological tools themselves are insufficient for efficient teaching strategies, which should always be accompanied by appropriate pedagogical approaches as discussed in \cite{Bryan_2006}, \cite{Bufasi_2022}, and \cite{Faresta_2024}. Digital applets and software that explicitly compare GR and Newtonian orbits are less common, and they are needed as digital twins to aid teaching and research in orbital mechanics. Thus, we contribute to the community with our own-developed \textsc{gr-orbit-toolkit}.\par

In this paper, we thus present a new Python-based package that solves a non-linear ODE derived from GR by following a post-Newtonian approach, as well as its Newtonian counterpart by using numerical algorithms. The first version (\textsc{v1.0.0}) of the software is hosted in our group's GitHub repository\footnote{\textsc{gr-orbit-toolkit:} https://github.com/cPhysPlus/gr-orbit-toolkit} and contains extensive documentation for users and developers who wish to utilize it and contribute to its development. The code follows the \texttt{PEP 8} guidelines for Python developers and is modular and highly portable. The base code is open source and released under the MIT license, so it can be easily expanded and seeks to become a community project. Our paper is organized as follows: in Section~\ref{sec:math_back} we present the target dynamical ODEs, in Section~\ref{sec:soft_deve} we explain how the code is structured, in Section~\ref{sec:simu_resu} we show examples of simulation outputs, and in Section~\ref{sec:conc} we summarize our findings and provide an overview of future code developments.

\section{Mathematical Background}\label{sec:math_back}
The two-body system in our simulation consists of an object orbiting around a massive body located at the origin of a 2D Cartesian coordinate system. To solve for the dynamics of this system, we have considered classical (Newtonian theory) and general relativistic (post-Newtonian approximation) mechanics.\par

\subsection{Classical Mechanics}\label{sec:cm_section}

In classical mechanics, the two-body problem is analytically solved by using the law of universal gravitation for the two interacting masses. This results in the relative motion equation, in which a test particle moves around a fixed massive center, similar to a one-body system. The solution to this equation is expressed through Kepler's laws of motion, which describe how objects follow orbits in the shape of conic curves \cite{Boccaletti1996}. Despite its simplicity, the classical approach does not apply to strong gravitational fields or objects moving at very high speeds (close to the speed of light). It also fails to explain certain observables, such as gravitational waves and effects like time dilation and length contraction \cite{Schutz_2009}.\par

The equation of motion in Newtonian theory is given by \cite{Hobson_2006}:
\begin{equation}\label{eq:ODE-classical}
    \frac{d^{2}u}{d\phi^{2}} + u = \frac{G\,M}{h^{2}},
\end{equation}
where $u = 1/r$ is the inverse of the radial distance from the central object, $\phi$ is the azimuthal angle, $h = r^{2} \, \dot{\phi}$ is the specific angular momentum of the orbiting body, $G$ is the universal gravitational constant, $M$ is the central object mass, and $c$ is the light speed in vacuum. For a bound orbit, the solution reads \cite{Hobson_2006}:
\begin{equation}\label{eq:solution-classical}
    u = \frac{G\,M}{h^{2}} \, (1 + e\,\cos{\phi}),
\end{equation}
where $e$ is the eccentricity of the orbit. For $0<e<1$, this solution represents an ellipse, where the closest distance between bodies (perihelion) is given by $a\,(1-e)$ and the furthest distance between bodies (aphelion) is given by $a\,(1+e)$. Here, $a$ is the semi-major axis and is defined as \cite{Hobson_2006}: $a = h^{2}/(G\,M\,(1-e^{2}))$.\par

Equation~\ref{eq:ODE-classical} is defined in terms of $u$ and $\phi$. To express it in terms of $r$ and $t$, we use the chain rule and the substitutions $\dot{\phi} = h\,u^2$ and $\frac{dr}{du}=-\frac{1}{u^2}$. The radial derivatives become: $\frac{dr}{dt} = -h \, \frac{du}{d\phi}$ and $\frac{d^{2}r}{dt^{2}} = -h^{2} \, u^{2} \, \frac{d^{2}u}{d\phi^{2}}$. Using these expressions, the radial acceleration $a_r$, including the centripetal term $-r\,\dot{\phi}^2$, becomes:
\begin{equation}\label{eq:radial_acceleration}
    \frac{dv}{dt} = a_r = \frac{d^{2}r}{dt^{2}} - r\,\dot{\phi}^2 \:\Rightarrow\: \frac{dv}{dt} = a_r = -\frac{h^2}{r^2} \left( \frac{d^2u}{d\phi^2} + u \right).
\end{equation}

\noindent Substituting Equation~\ref{eq:ODE-classical} into Equation~\ref{eq:radial_acceleration}, we recover the inverse-square law for radial acceleration and Newton's gravitational force in vector form:
\begin{equation}\label{eq:classical-corrected-acc}
    \frac{dv}{dt} = a_r = -\frac{G\,M}{r^2} \Rightarrow \boxed{    \vec{F}_{\text{Newton}} = -\frac{G\,M\,m}{r^3} \, \vec{r}.}
\end{equation}

\noindent We can express Equation \ref{eq:classical-corrected-acc} as the following system of ODEs, in vector form:
\begin{equation}
\boxed{
\frac{d\vec{r}}{dt} = \vec{v}, \quad \frac{d\vec{v}}{dt} = -\frac{GM}{r^3} \vec{r}.} \label{eq:coupled-ODEs-classical}
\end{equation}

From Equation~\ref{eq:ODE-classical}, we can also derive Kepler's laws. Particularly, by considering angular momentum conservation and integrating the equation over a full period $T$, we obtain the third law, which reads \cite{Goldstein_2001}: $ 4\,\pi^{2}\,a^{3} = G\,M\,T^{2}$.

\subsection{General Relativistic Mechanics}\label{sec:gr_section}

In GR, the two-body problem is solved in the realm of Schwarzschild geometry, which describes the space-time outside a non-rotating, spherically symmetric mass \cite{Bambi_2018}. Particularly, in the post-Newtonian approximation, GR corrections are added to the classical law of universal gravitation. This process begins with a Hamiltonian, to which we add correction terms of a specific order. We then solve the equations of motion in a canonical space and transform them back into Keplerian-like terms \cite{Blanchet_2011}. While this is an approximation, the method allows us to account for relativistic effects in the presence of strong gravitational fields.\par

The equation of motion in GR is given by \cite{Hobson_2006}:
\begin{equation}\label{eq:ODE-relativistic}
    \frac{d^{2}u}{d\phi^{2}} + u = \frac{G\,M}{h^{2}} + \frac{3\,G\,M}{c^{2}}\,u^{2}.
\end{equation}

This equation does not have an analytical solution. One method for approximating it involves perturbation theory, where the Newtonian solution (Equation~\ref{eq:solution-classical}) corresponds to its zeroth-order solution, and a higher-order perturbation term is added to be solved iteratively. With this and defining $\alpha = 3\,(G\,M)^{2} / (h^{2}\,c^{2}) \ll 1$, we obtain the following approximated solution \cite{Hobson_2006}:
\begin{equation}
    u \approx \frac{G\,M}{h^{2}} \, \{ 1 + e\,\cos{[\phi\,(1 - \alpha)]}\}.
\end{equation}

This solution translates into the precession of the elliptical orbit by an amount $\Delta\phi$ at each rotation, given by \cite{Hobson_2006}:
\begin{equation}\label{eq:precession}
    \Delta\phi = \frac{6\,\pi\,G\,M}{a\,(1-e^{2})\,c^{2}}.
\end{equation}

Another essential quantity in GR is the Schwarzschild radius ($r_{s}$), which is the radius of a sphere that marks the limit of a massive object's radius before becoming a black hole when compressed to that size. This quantity is \cite{Kogut_2018}:
\begin{equation}\label{schw_radius}
    r_{s} = \frac{2\,G\,M}{c^{2}}
\end{equation}

Following the same steps as in the classical derivation, the radial acceleration retains the form of Equation~\ref{eq:radial_acceleration}. Substituting Equation~\ref{eq:ODE-relativistic} in Equation~\ref{eq:radial_acceleration} yields the GR-corrected acceleration (see also \cite{D’Eliseo_2007}) and the corresponding GR force:
\begin{equation}\label{eq:relativistic-corrected-acc}
    \frac{dv}{dt} = a_r = -\frac{G\,M}{r^2} \left( 1 + \frac{3\,h^2}{c^2\,r^2} \right)\:\Rightarrow\:\boxed{ \vec{F}_{\text{GR}} = -\frac{G\,M\,m}{r^3} \, \vec{r} \left( 1 + \frac{3\,h^2}{c^2\,r^2} \right).}
\end{equation}

\noindent Notice that for circular orbits $h^2 = G\,M\,r$, so the dominant relativistic correction in $\ddot{r}$ is: $-3\,G\,M\,h^2/(c^2\,r^4)$. We can express Equation \ref{eq:relativistic-corrected-acc} as the following system of ODEs, in vector form:
\begin{equation}
\boxed{
\frac{d\vec{r}}{dt} = \vec{v}, \quad \frac{d\vec{v}}{dt} = -\frac{GM}{r^3} \vec{r} \left( 1 + \frac{3h^2}{c^2 r^2} \right).} \label{eq:coupled-ODEs-relativistic}
\end{equation}

\section{\textsc{gr-orbit-toolkit}: Software Development}\label{sec:soft_deve}

\subsection{Code Structure}\label{sec:toolkit}

The \textsc{gr-orbit-toolkit} is a Python-based package for simulating the motion of a small body (e.g., an Earth-sized planet) around a massive object (e.g., the Sun or a black hole) by incorporating both classical and relativistic dynamics. The code follows \texttt{PEP 8} style guidelines\footnote{https://peps.python.org/pep-0008}. It accepts user-defined parameters (see Table~\ref{tab:parameters} for default values), numerically solves the classical and relativistic acceleration ODEs (Equations~\ref{eq:coupled-ODEs-classical} and \ref{eq:coupled-ODEs-relativistic}, respectively), and outputs orbital histories, animations, precomputed grids, and comparison plots. The toolkit is modular, allowing extensions for other physical systems.\par

\begin{table}
\centering
\caption{Default input parameters for the \textsc{gr-orbit-toolkit}.}\label{tab:parameters}
\begin{tabular}{|l|l|l|l|}
\hline
\textbf{Input} & \textbf{Type} & \textbf{Description} & \textbf{Default}\\
\hline
\texttt{ecc} & \texttt{float} & Eccentricity of the orbit & $0.0$\\
\texttt{mass\_bh} & \texttt{float} & Mass of the central object in solar masses [$M_\odot$] & $5.0 \rm e6$\\
\texttt{sm\_axis} & \texttt{float} & Semi-major axis of the orbit in astronomical units [AU] & $1.0$\\
\texttt{orb\_period} & \texttt{float} & Number of orbital periods to simulate & $2.0$\\
\texttt{method} & \texttt{str} & Numerical method for ODE integration & `RK3'\\
\texttt{relativity} & \texttt{bool} & Whether or not to include relativistic effects & `False'\\
\hline
\end{tabular}
\end{table}

The core module (\texttt{orbits.py}) is structured into four classes: 

\begin{enumerate}
    \item \texttt{TwoBodyProblem} computes the classical/relativistic orbital dynamics and generates the initial conditions and grids.
    \item \texttt{SimulationRunner} executes the simulation using the selected numerical method and saves the orbital histories in output files.
    \item \texttt{AnalysisTools} compares simulation results (classical versus relativistic and different integration methods) and analyzes convergence and errors.
    \item \texttt{AnimationCreator} generates orbit animations from simulation data for visualization and analysis.
\end{enumerate}

\begin{figure}
    \centering
    \makebox[\textwidth]{\includegraphics[width=0.88\linewidth]{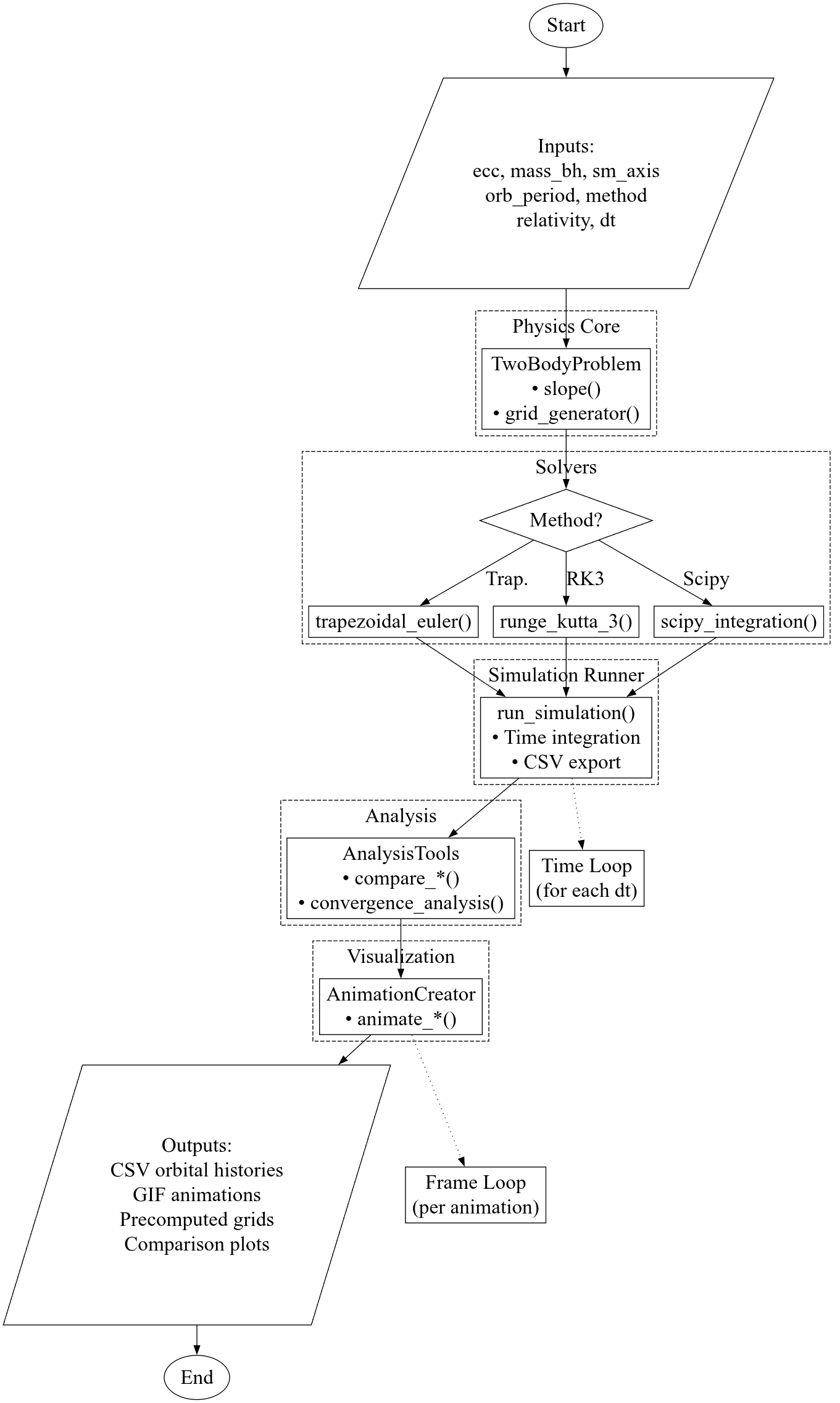}}
    \caption{Workflow diagram of the orbital simulation module \texttt{orbits.py} for \textsc{gr-orbit-toolkit}. Created using Graphviz Online \cite{Graphviz}.}
    \label{fig:workflow}
\end{figure}

Figure~\ref{fig:workflow} illustrates the structure of the modules as a workflow diagram.

\subsection{Numerical ODE Integrators}\label{sec:numerics}

The acceleration ODEs (Equations~\ref{eq:coupled-ODEs-classical} and \ref{eq:coupled-ODEs-relativistic}) are solved numerically using one of the following integration methods, selectable by the user:

\begin{itemize}
\item The Trapezoidal Euler method is a predictor-corrector approach that utilizes the explicit Euler method as the predictor step, the implicit Euler method as the corrector step, and the trapezoidal rule for averaging, resulting in a second-order accurate method. Our numerical implementation follows \cite{Kong_2021}:
\begin{equation}
    \begin{aligned}
        v_{n+1} &= v_{n} + \Delta t \cdot a\left(r_{n}, v_{n}\right) \\
        v_{n+1} &= v_{n} + \frac{\Delta t}{2} \cdot \left[a\left(r_{n}, v_{n}\right), v_{n}) + a\left(r_{n+1}, v_{n+1}\right)\right].
    \end{aligned}
\end{equation}

\item The Runge-Kutta 3 (RK3) method is a third-order, explicit scheme that uses weighted slopes ($k_{1}$, $k_{2}$, $k_{3}$) for better accuracy compared to an Euler method \cite{Kong_2021}. Our implementation uses a modified weighting of the coefficients based on the Butcher tableau technique \cite{Butcher_2016}:
\begin{equation}
    \begin{aligned}
        k_{1} &= a\left(r_{n}, v_{n}\right) \\
        k_{2} &= a\left(r_{n} + \frac{3}{4} \cdot \Delta t, v_{n} + \frac{3}{4} \cdot \Delta t \cdot k_{1}\right) \\
        k_{3} &= a\left(r_{n} + \frac{1}{4} \cdot \Delta t, v_{n} - \frac{5}{12} \cdot \Delta t \cdot k_{1} + \frac{2}{3} \cdot \Delta t \cdot k_{2}\right) \\
        v_{n+1} &= v_{n} + \frac{\Delta t}{9} \cdot \left(k_{1} + 5 \cdot k_{2} + 3 \cdot k_{3}\right).
    \end{aligned}
\end{equation}
\item SciPy's Dormand \& Prince (hence DOP853) method is an explicit eight-order Runge-Kutta method with error tolerance control \cite{Hairer_1993}. We use DOP853 as the reference method for numerical comparisons in Section \ref{subsec:numerics}. If desired, \textsc{gr-orbit-toolkit} code developers can readily write additional ODE integrators as class methods within our \texttt{TwoBodyProblem} class.\par
\end{itemize}

\subsection{Command-Line Interface and Computing Requirements}\label{sec:cli}
The simulation toolkit employs Python's \texttt{argparse} library to provide a flexible command-line interface (CLI) for user customisation. The interface organizes input options into groups: Simulation Parameters (e.g., \texttt{-\!-ecc 0.01}), Analysis Options (e.g., \texttt{-\!-compare\_rel\_class}), and Output Options (e.g., \texttt{-\!-save\_gif}). The help text (\texttt{-h/-\!-help}) automatically lists all the parameters with type hints and default values. To view all the available flags, a description, and the default values, check the Available Flags section of the \texttt{README.md} file in our code's repository. If no flags are specified, the simulation runs with the default values (see Table \ref{tab:parameters}).\par

In terms of computing resources, the \textsc{gr-orbit-toolkit} is inexpensive. The code runs in serial mode on CPU architectures. To illustrate, the simulations presented in the next section were executed in single-core mode on a CPU-based, portable computer with a clock speed frequency of $2.30$ GHz and $10$ GB of Random Access Memory (RAM). For a representative test case with the default parameters (200-time steps per orbital period), a complete simulation comparing classical and relativistic approaches required $\sim$ 90 seconds to run.\par

\section{Software Usage and Simulation Results}\label{sec:simu_resu}

\subsection{Simulating Orbits with \textsc{gr-orbit-toolkit}}
The \textsc{gr-orbit-toolkit} is publicly available under the MIT license. Installation requires Python 3.9 or later and can be completed by: 1) \texttt{git}\footnote{https://git-scm.com} cloning the GitHub repository hosting the package, and 2) installing in editable mode with \texttt{pip}\footnote{https://pypi.org/project/pip}. We provide a complete guide for users and developers on using the package in the \texttt{README.md} file of the repository. In this paper, we report 7 simulations (see Table~\ref{tab:reported-simulations}). A sample user-customized CLI execution statement of the first model in the table is:\par
\begin{verbatim}
python -m orbits --ecc 0.01671 --mass_bh 1. --orb_period 1.\
       --compare_rel_class --save_gif   
\end{verbatim}

\begin{table}
\centering
\caption{Simulations reported in this paper.}\label{tab:reported-simulations}
\begin{tabular}{|l|l|l|l|}
\hline
\textbf{Identifier} & \textbf{System} & \textbf{Eccentricity} & \textbf{Method} \\
\hline
1 & Earth-Sun & 0.01671 & RK3 \\
2 & Earth-black hole & 0.0 & RK3 \\
3 & Earth-black hole & 0.01671 & RK3 \\
4 & Pluto-black hole & 0.25 & Trapezoidal \\
5 & 7092 Cadmus-black hole & 0.70 & Trapezoidal \\
6 & Earth-black hole & 0.01671 & Trapezoidal \\
7 & Earth-black hole & 0.01671 & SciPy \\
\hline
\end{tabular}
\end{table}

\begin{figure}[!ht]
\begin{center}
  \begin{tabular}{c c}
     \hspace{-1.5cm}\resizebox{75mm}{!}{\includegraphics[height=7.5cm]{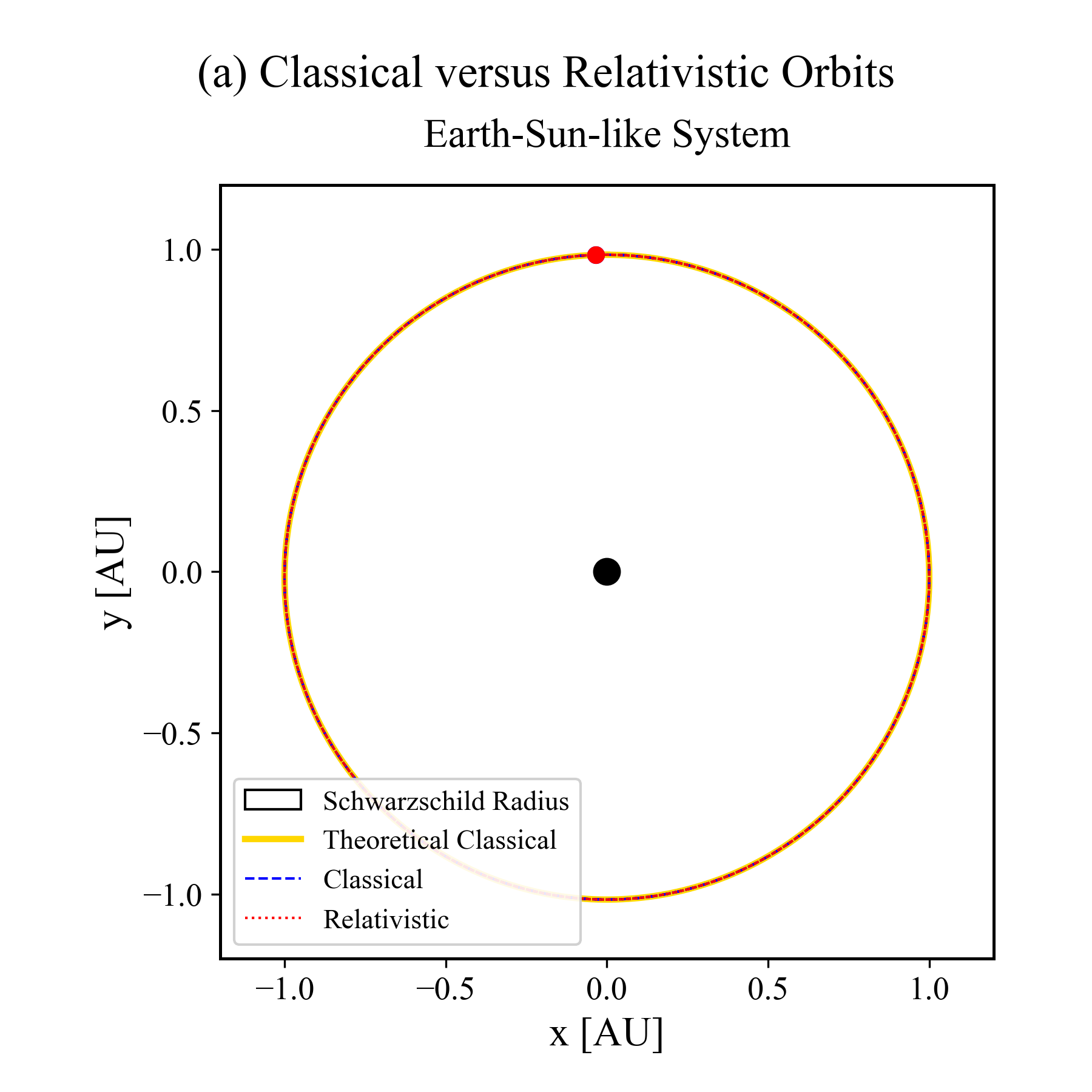}} & \hspace{-0.3cm}\resizebox{75mm}{!}{\includegraphics[height=7.5cm]{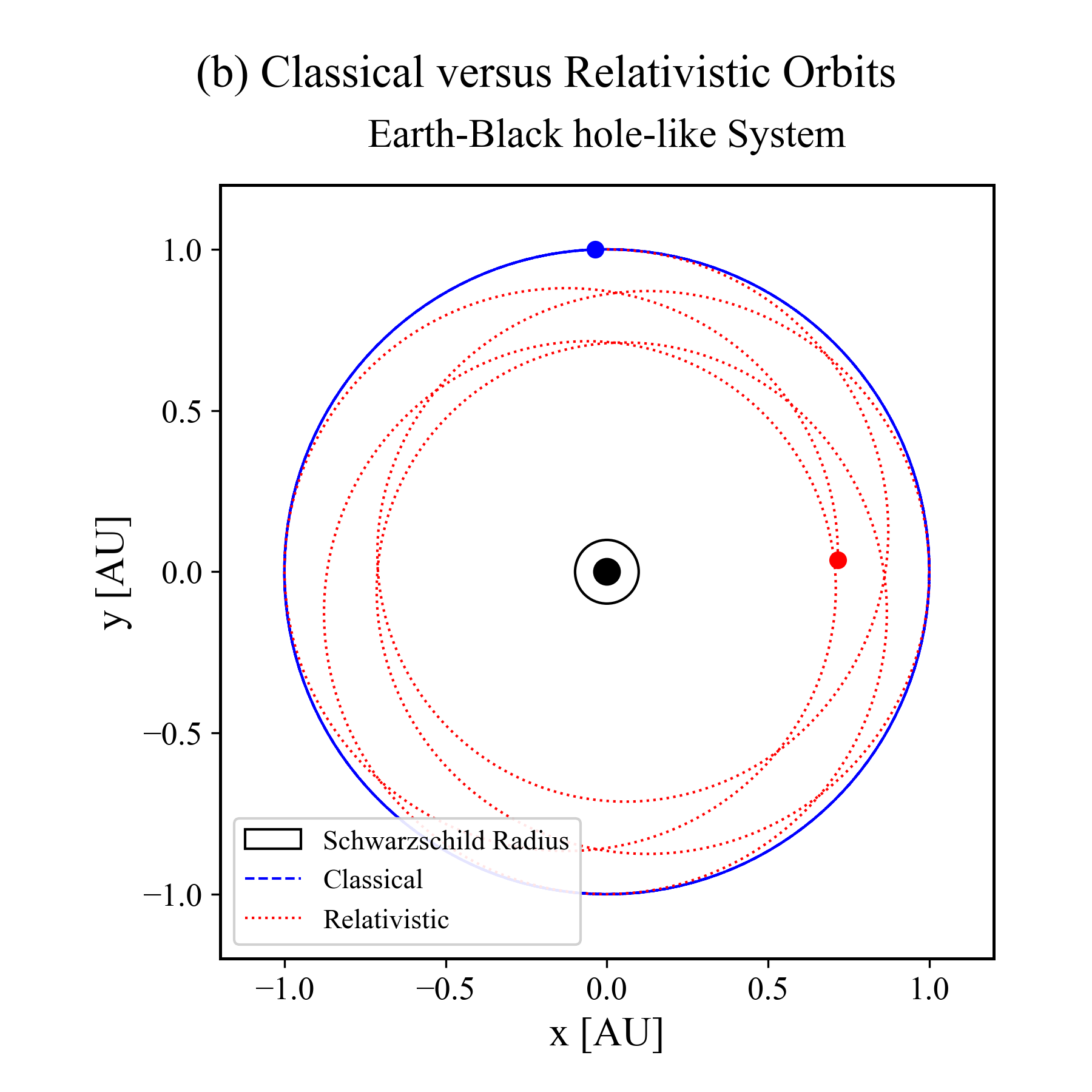}}\\
  \end{tabular}
  \caption{Classical (blue dashed line) versus relativistic (red dotted line) orbits of two systems: (a) an Earth-Sun-like system ($e=0.01671$) and (b) an Earth-black hole-like system ($e=0.0$). All models were run with the RK3 method. The theoretical classical orbit (yellow solid line) is shown for reference. For central objects with little mass, such as the Sun (a), classical and relativistic orbits are indistinguishable, while for supermassive central objects (b), such as a black hole, relativistic precession is observed.}
  \label{fig:class_vs_relat}
\end{center}
\end{figure}

In this execution example, we consider an Earth-sized planet orbiting around the Sun (of Schwarzschild radius $r_{s} = 2.953 \cdot 10^{3}$ [m]). The \textsc{gr-orbit-toolkit} generates multiple output files, including orbital histories in \texttt{CSV} format, \texttt{GIF} animations, and precomputed grids in \texttt{.npy} (standard binary file) format. All these files are automatically stored in the \texttt{outputfolder} within the working directory. At the same time, the comparison plots are displayed once the simulation is run if one of the corresponding flags is used. The above CLI execution line also generates an animation comparing relativistic and classical orbits in the system composed of the Earth (with an eccentricity of 0.01671) orbiting the Sun (of 1.0 $M_\odot$). Figure~\ref{fig:class_vs_relat}(a) shows the last frame of this animation, in addition to a comparison between our numerical results and the theoretical prediction from Equation~\ref{eq:solution-classical}. As we explored some existing simulation applets, we found that there is no simple way to extract the orbital data needed for comparisons, but the agreement between the analytical and numerical orbits validates our code. The reader can find the full GIF animations of the above and following simulations in the \texttt{examples} folder in the public code repository.\par 

\vspace{-0.7cm}
\vspace{-0.7cm}

Figure~\ref{fig:class_vs_relat}(a) shows the integrated classical and relativistic orbital histories of an Earth-sized planet around the Sun after one orbital period. For all the simulations, we considered the following initial conditions (position and velocity, respectively): $(x_{0},y_{0}) = (0,a\,(1-e))$ and $(v_{x0},v_{y0}) = (-\sqrt{(G\,M/a\,(1+e)/(1-e)},0)$. As expected, both orbits display almost perfect circular trajectories. The classical and relativistic orbits exhibit negligible differences in this set of simulations because the mass of the Sun is not high enough to make relativistic effects (e.g., orbital precession) significant. The eccentricity of the Earth-sized planet is also close to circular, which prevents the body from approaching the massive object and experiencing strong relativistic forces.\par

\subsection{Relativistic Precession and the Role of Eccentricity}

Figure~\ref{fig:class_vs_relat}(b) shows the last frame snapshot of the simulation comparing classical versus relativistic orbits for a system composed of an Earth-sized planet orbiting around a black hole with $5\times10^6\,M_\odot$ (of Schwarzschild radius $r_{s} = 1.477 \cdot 10^{10}$ [m]). This and subsequent simulations were run using the default values for the input parameters (Table~\ref{tab:parameters}), except for the orbital period. With the change in the values of eccentricity (to zero) and the mass of the central object (to depict a supermassive black hole), we observe a remarkable difference between classical and relativistic orbits for the same system and identical initial conditions. The classical orbit follows a perfectly circular path, while the relativistic orbit follows a different path that precesses by $0.930$ [rad] per orbit, as dictated by Equation~\ref{eq:precession}. This result confirms that classical mechanics is not an appropriate theory to describe the orbital motion of objects orbiting massive bodies, such as black holes. It also indicates that our code in the \textsc{gr-orbit-toolkit} and its numerical framework correctly capture relativistic precession \cite{Marin_2018}. Our simulations also confirm that the orbital motions predicted by classical and relativistic ODEs drastically differ for bodies near the Schwarzschild radius of the central massive body, which is substantially ($\sim 7\,\rm dex.$) larger for the black hole.\par

\begin{figure}[!ht]
\begin{center}
  \begin{tabular}{c c}
     \hspace{-1.5cm}\resizebox{75mm}{!}{\includegraphics[height=7.5cm]{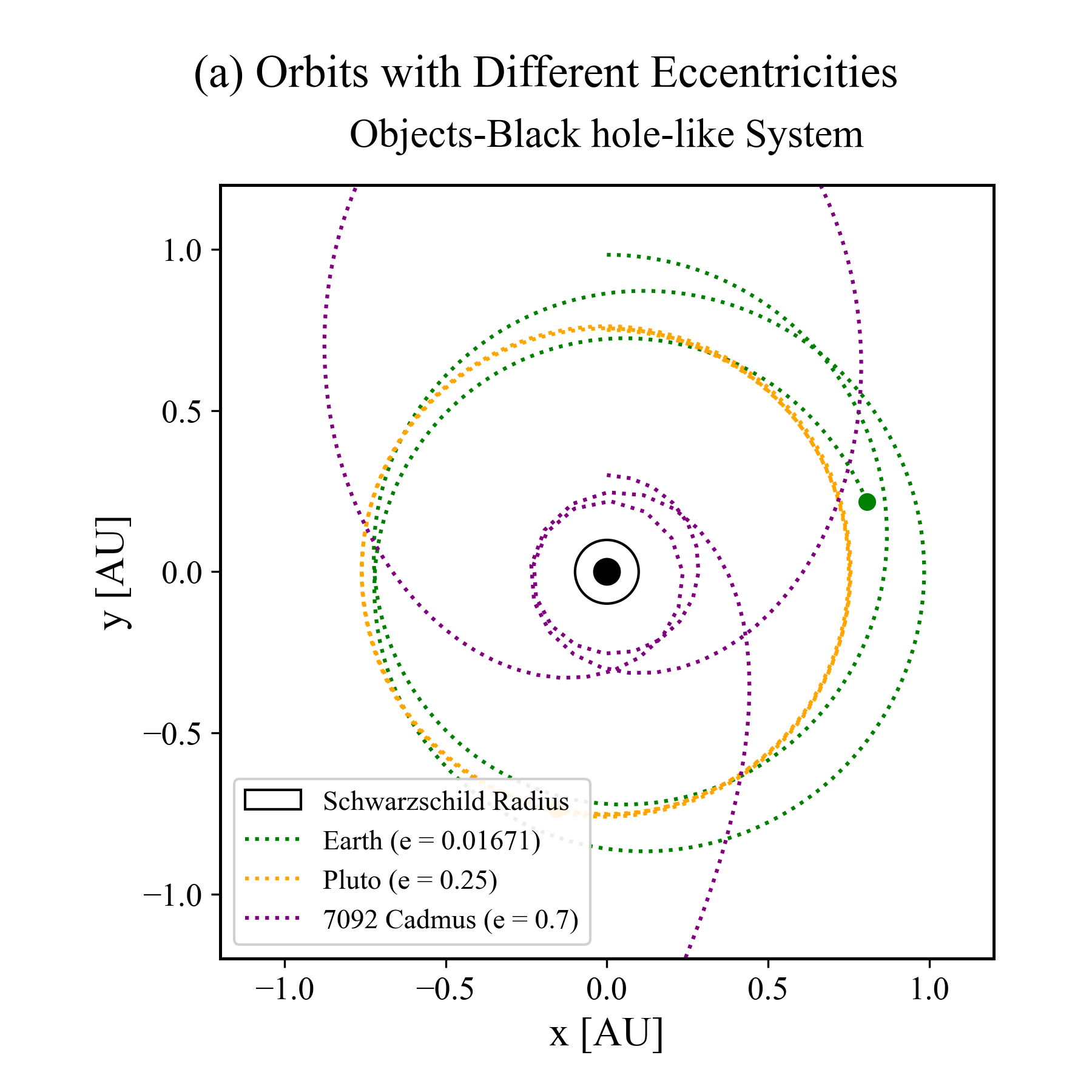}} & \hspace{-0.3cm}\resizebox{75mm}{!}{\includegraphics[height=7.5cm]{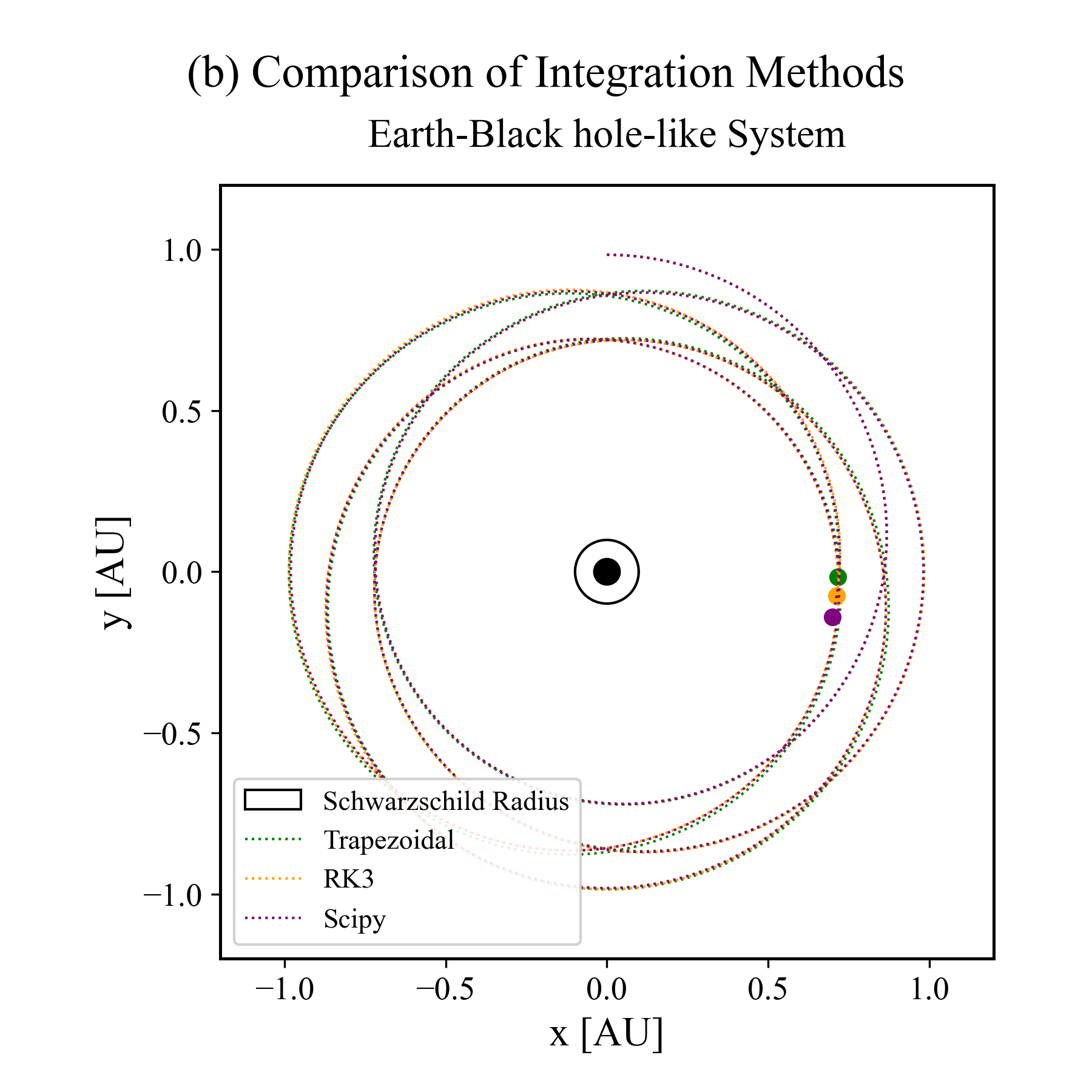}}\\
  \end{tabular}
  \caption{(a) Relativistic orbits of various objects with different eccentricities surrounding a black hole, run with Trapezoidal integration. (b) Relativistic orbits of an Earth-like planet-black hole system with $e=0.01671$, using several ODE integration methods. Objects with lower eccentricities follow more circular paths, and higher eccentricities dictate a more elliptical path rotation. The Trapezoidal and RK3 integration methods slightly underestimate the orbit, but are consistent with SciPy's higher-order method.} \label{fig:eccs_methods}
\end{center}
\end{figure}

We also studied the role of eccentricity. Figure~\ref{fig:eccs_methods}(a) shows the last frame snapshot of the simulations comparing the orbits of three objects with different eccentricity values (see Table \ref{tab:reported-simulations}). Apart from observing relativistic precession, we also see that objects with lower eccentricities follow more circular trajectories. In comparison, objects with higher initial eccentricities follow more elliptical or open trajectories. The case of 7092 Cadmus is particularly interesting as the object completes a couple of orbits around the black hole, but then it slingshots outside of the inner $1$-AU region of the system. These simulations are, of course, idealized, but they highlight the importance of incorporating relativistic corrections into the equations of motion for bodies orbiting near massive objects, such as black holes. These results also show that despite the approximations made in the post-Newtonian treatment of the relativistic equations of motion (see Section 
\ref{sec:math_back}), our toolkit can be effectively used to visualize relativistic effects on orbits. Due to its user-oriented framework, in-built input/output and visualization tools, and flexibility, the \textsc{gr-orbit-toolkit} can be used as a digital twin simulation package in STEM education. In future work, we plan to study its educational effectiveness through classroom deployments and user surveys.\par

\subsection{Numerical Convergence}
\label{subsec:numerics}
\begin{figure}[!ht]
\centering
\includegraphics[height=7.5cm]{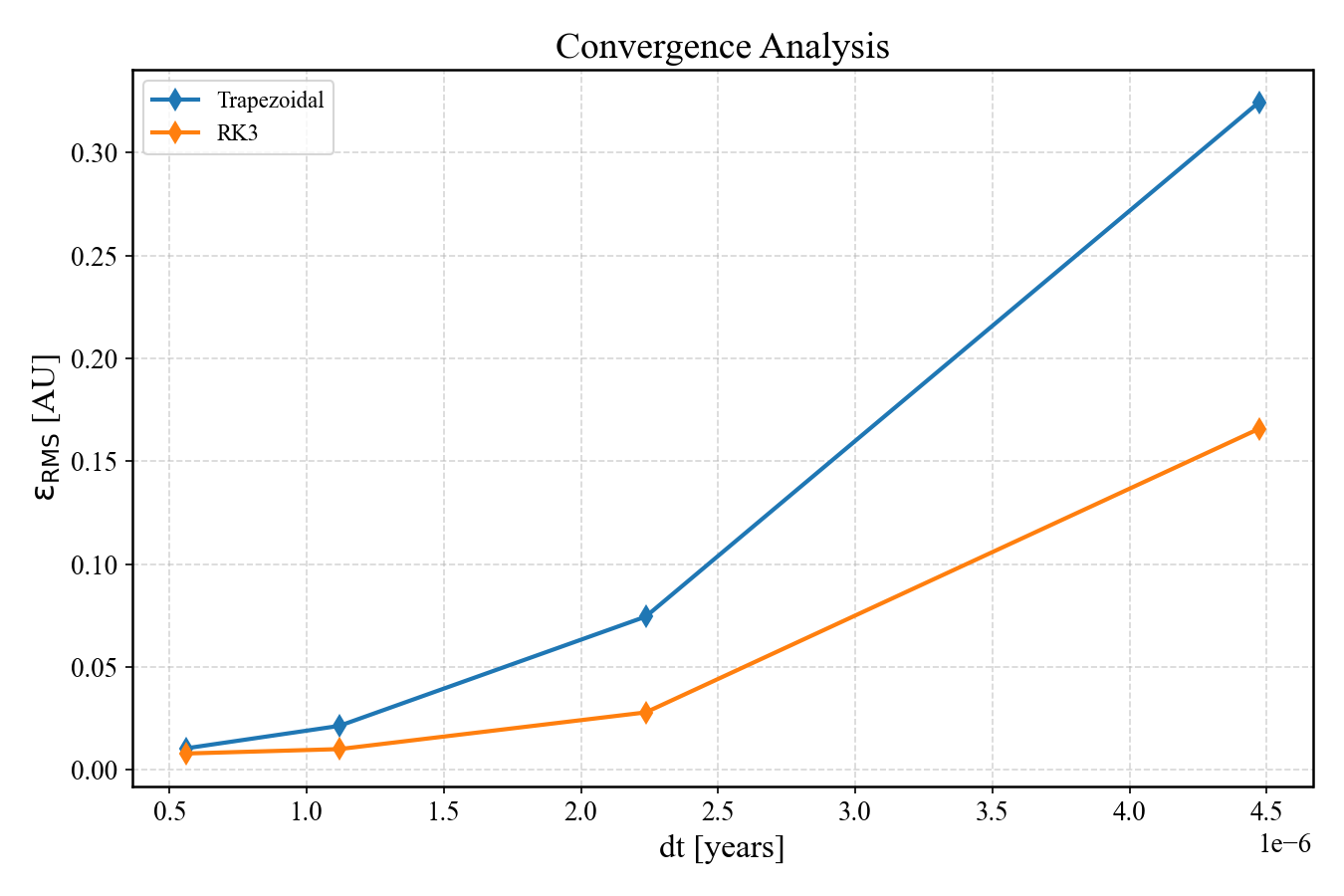}
\caption{Convergence analysis showing the RMS error of the Trapezoidal and RK3 schemes with respect to the reference method (SciPy's DOP853) for several time steps. The RMS error increases as the time step becomes larger. For all the considered time steps (time resolutions), the RK3 method is more accurate than the Trapezoidal one.}
\label{fig:convergence}
\end{figure}

Figure~\ref{fig:eccs_methods}(b) shows the last frame snapshot of the simulations (which corresponds to the 600th time step or 3 orbital periods) comparing the orbits followed by a standard Earth-black hole system, but integrated with all three different integration methods currently available in the \textsc{gr-orbit-toolkit}. We then plot the root mean square (RMS) error (Figure~\ref{fig:convergence}) between the numerically defined methods (Trapezoidal Euler and RK3) and a reference method (SciPy's DOP853) obtained by integrating with several time steps, with a value of $0.01671$ for the eccentricity.\par

The snapshot (Figure~\ref{fig:eccs_methods}(b)) shows the paths followed by the same object when the relativistic ODEs are integrated with different integration methods. We observe that, compared to the eight-order SciPy method, the Trapezoidal and RK3 methods slightly underestimate the orbit trajectory. This difference is quantitatively defined in the plot comparing the RMS error between methods (Figure~\ref{fig:convergence}). Here, we also observe that the Trapezoidal-SciPy RMS error is higher than the RK3-SciPy RMS error for all the time steps we considered. Notice that as the time step ($\rm dt$) decreases, the curves tend to flatten and appear to reach a plateau. This indicates the numerical solutions reach machine precision at those time resolutions. Overall, these results demonstrate that the \textsc{gr-orbit-toolkit} is numerically robust.\par
\vspace{-0.7cm}

\vspace{-0.7cm}

\section{Conclusions}\label{sec:conc}

We have presented \textsc{gr-orbit-toolkit}, an open-source software for simulating the dynamics of a two-body system using both classical and relativistic approaches. By combining user-friendly, command-line parameters with real-time visualization and analysis tools, the toolkit proves to be a valuable simulation tool for orbital motion. The main features of our software are:\par

\begin{itemize}
    \item The \textsc{gr-orbit-toolkit} is modular and structured in Python classes. The code design follows \texttt{PEP 8} guidelines and aims to become a digital twin for STEM education in physics, with particular attention to general relativity and orbital mechanics.
    \item The toolkit can be readily customized by users and expanded by developers, for whom we provide an extensive guide in the code repository. Code execution requires a single-line CLI statement where user inputs are passed to the simulation class via \texttt{argparse} flags.
    \item Our software incorporates three ODE integrators (Euler Trapezoidal, RK3, and SciPy's DOP853), which the user can select to solve either classical or relativistic ODEs. The acceleration ODEs are derived from standard Newtonian (classical) and post-Newtonian (relativistic) approaches.
\end{itemize}

\noindent We have also reported a set of simulations that demonstrate the capabilities of the \textsc{gr-orbit-toolkit} to model realistic physical systems, ranging from planets orbiting our Sun to objects with different eccentricities orbiting a central black hole, where relativistic corrections can be significant. Our main conclusions are: 

\begin{itemize}
    \item For an Earth-Sun-like system, the difference between the classical and relativistic orbits is negligible. As expected, the orbits follow almost perfect circular trajectories due to the Sun's low mass and the planet's low eccentricity. No relativistic effects are observed in the Earth-Sun-like system.
    \item For a system with an Earth-sized object orbiting a supermassive black hole, relativistic effects are noticeable. Classical orbits follow circular paths, while relativistic orbits present trajectories that precess at each rotation. Our code adequately captures relativistic precession.
    \item Our software can also simulate systems of objects with different eccentricities orbiting around the central body. We have studied three cases, finding that high eccentricities change the path of the orbit from more circular to more elliptical or open orbits while featuring precession.
    \item By keeping the values of eccentricity and central mass fixed and varying the ODEs integration methods used for simulating a relativistic system, we observe that the RK3 method is more accurate than the Trapezoidal method when compared to the higher-order SciPy's DOP853 method. Overall, all methods show consistent numerical results.
\end{itemize}

\noindent While our toolkit successfully simulates individual two-body systems, it also has limitations. For instance, the adopted relativistic (post-Newtonian) approximation only captures leading-order relativistic effects, neglecting higher-order terms \cite{Blanchet_2011}. Therefore, extending the algorithm to consider more general equations and additional bodies is a natural next step. Similarly, the currently available single-CPU implementation may limit the number of simulations required to create a relativistic orbit catalog. Future versions could also include code parallelization options and develop memory-usage predictors, which this version does not have. With its modular architecture, we aim for this toolkit to become a collaborative project in computational physics. Evaluations of its effectiveness as an educational tool in classrooms will be conducted in the future. The public code repository on GitHub provides interactive notebooks to guide users.\par

\subsubsection{Acknowledgments}
We thank the anonymous referees for their insightful comments that helped improve this manuscript. The authors acknowledge CEDIA (\url{www.cedia.edu.ec}) for providing access to its high performance computing (HPC) cluster, which was essential to develop the software presented in this paper.


\bibliographystyle{splncs04}
\bibliography{mybibliography}

\end{document}